# An Ultralow Leakage Synaptic Scaling Homeostatic Plasticity Circuit With Configurable Time Scales up to 100 ks

Ning Qiao, *Member, IEEE*, Chiara Bartolozzi, *Member, IEEE*, and Giacomo Indiveri, *Senior Member, IEEE*

*Abstract*—Homeostatic plasticity is a stabilizing mechanism commonly observed in real neural systems that allows neurons to maintain their activity around a functional operating point. This phenomenon can be used in neuromorphic systems to compensate for slowly changing conditions or chronic shifts in the system configuration. However, to avoid interference with other adaptation or learning processes active in the neuromorphic system, it is important that the homeostatic plasticity mechanism operates on time scales that are much longer than conventional synaptic plasticity ones. In this paper we present an ultralow leakage circuit, integrated into an automatic gain control scheme, that can implement the synaptic scaling homeostatic process over extremely long time scales. Synaptic scaling consists in globally scaling the synaptic weights of all synapses impinging onto a neuron maintaining their relative differences, to preserve the effects of learning. The scheme we propose controls the global gain of analog log-domain synapse circuits to keep the neuron's average firing rate constant around a set operating point, over extremely long time scales. To validate the proposed scheme, we implemented the ultralow leakage synaptic scaling homeostatic plasticity circuit in a standard 0.18 $\mu$m complementary metal-oxide-semiconductor process, and integrated it in an array of dynamic synapses connected to an adaptive integrate and fire neuron. The circuit occupies a silicon area of 84 $\mu$m × 22 $\mu$m and consumes approximately 10.8 nW with a 1.8 V supply voltage. We present experimental results from the homeostatic circuit and demonstrate how it can be configured to exhibit time scales of up to 100 ks, thanks to a controllable leakage current that can be scaled down to 0.45 aA (2.8 electrons per second).

## I. Introduction

ONE of the most remarkable properties of nervous systems is their ability of adapting to the changes in the environment, in order to achieve and maintain robust neural computation. This ability is mediated by multiple forms of *plasticity*, acting on a wide range of different time scales [1]. The modification of synaptic weights over very short temporal windows (i.e., in the order of milliseconds) is believed to attain selectivity to transient stimuli and contrast adaptation [2]. Post-synaptic long term plasticity mechanisms (e.g., that settle in tens to hundreds of milliseconds) such as Spike-Timing Dependent Plasticity (STDP), mediate classical neural network learning processes [3]. Longer term changes in synaptic transmission and intrinsic excitability of the neurons (e.g., that settle over minutes to days) have been shown to mediate homeostatic control that keeps the neuron's activity within functional bounds [4].

In engineering terms, homeostatic plasticity is a form of automatic gain control that counteracts the effect of long lasting drifts of the neurons activity due to changes in external conditions, in input activity levels, due to temperature variations or to changes in internal connectivity. This mechanism is therefore extremely valuable for the effective deployment of hardware implementations of spiking neural networks, as it increases robustness to long-lasting changes in the operating conditions of the system by automatically tuning the network internal parameters. Despite its crucial role for the design of large scale spiking neural networks, only few works have been devoted to the implementation of homeostatic plasticity, mostly due to the difficulty in achieving the necessary long time constants on silicon. With the exception of [5], that acts on neural excitability, the work proposed so far focused on a specific form of synaptic plasticity, known as synaptic scaling, that modulates a neuron's activity by modifying its total synaptic drive [6]. This form of homeostatic control scales the strength of the synapses connected to a single neuron when its activity chronically changes. This global gain tuning preserves the relative differences between individual synapses, and does not disrupt the effect of activity dependent learning mediated by forms of Hebbian or Spike-Timing Dependent Plasticity mechanisms [7].

Long time scales in plasticity circuits have been obtained in the past thanks to the use of floating gate transistors [5], [8], or by resorting to the use of hybrid systems where the control is implemented in software with digital hardware in the loop, requiring external memory and conventional computing architectures [9]. In this work, we extend a previous proof-of-concept implementation [10] and show how it is possible to achieve extremely long time scales by exploiting an ultra-low

Manuscript received June 3, 2017; revised August 10, 2017; accepted September 10, 2017. This work was supported in part by EU ERC under Grant "NeuroP" (257219) and in part by EU ICT under Grant "NeuRAM$^3$" (687299). This paper was recommended by Associate Editor M. Delgado-Restituto. *(Corresponding author: Ning Qiao.)*
G. Indiveri and Q. Ning are with the Institute of Neuroinformatics, University of Zurich and ETH Zurich 8057, Switzerland (e-mail: giacomo@ini.uzh.ch; qiaoning@ini.uzh.ch).
C. Bartolozzi is with the Istituto Italiano di Tecnologia, Genova 16163, Italy (e-mail: Chiara.Bartolozzi@iit.it).
Color versions of one or more of the figures in this paper are available online at http://ieeexplore.ieee.org.
Digital Object Identifier 10.1109/TBCAS.2017.2754383

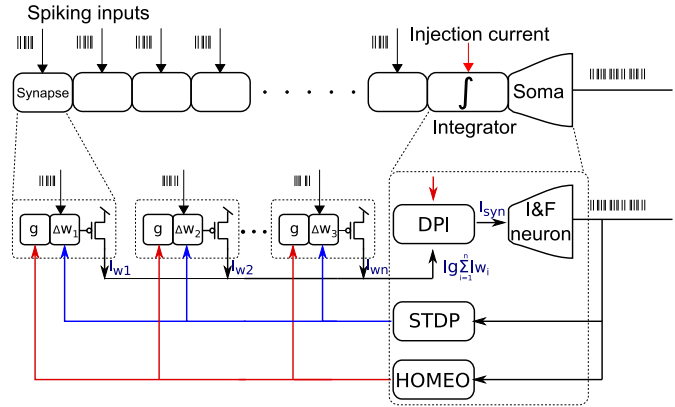

Fig. 1. Block diagram of a neuron with its input synaptic array. The DPI synapse is a linear integrator that adds input from different sources, each with two independent parameters for tuning the synaptic efficacy: "g" is used to globally scale of the synapses impinging of the same neuron, "$w_i$" is modified individually in each synapse by local short term and spike-based forms of plasticity. The homeostatic control loop reads out the total synaptic drive of the neuron, that in absence of current injected in the membrane capacitance corresponds to the total spiking activity of the neuron.

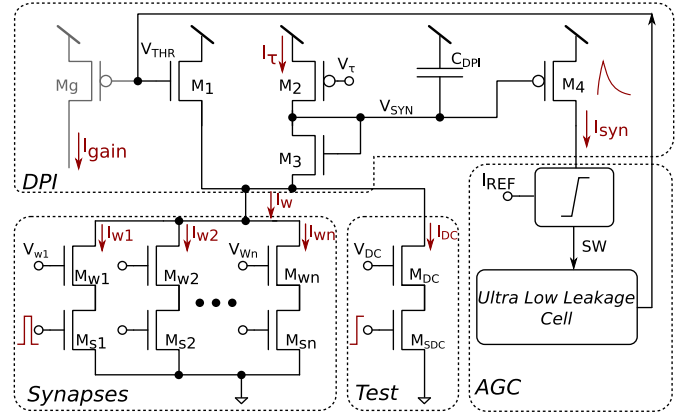

Fig. 2. Block diagram of proposed homeostatic AGC loop. The output current of the DPI block $I_{syn}$ is scaled automatically over long time scales, by up- or down-regulating the $V_{THR}$ control voltage. $M_{s1}$-$M_{sn}$ are switched by input synaptic pulses for generating synaptic pulse currents, while $M_{w1}$-$M_{wn}$ are gated by weight voltages $V_{w1}$-$V_{wn}$ in each synapses for particular weight current amplitude. A Test branch is added to supply a DC input current $I_{DC}$ with amplitude tuned by $V_{DC}$. It is applied as a synaptic input current into the circuit's DPI block to simplify the test. $M_g$ is a virtual PMOS gated by $V_{THR}$ for representing the current term $I_{\text{gain}}$ expressed in (2).

leakage cell [10]–[12] implemented on standard Complementary Metal-Oxide-Semiconductor (CMOS) technology within a novel auto-gain synaptic scaling circuit. The novelty associated with this work lies mainly in the new CMOS circuit design that can exhibit longer time scales and provide better control of the leakage currents. The synaptic scaling mechanism we propose is tailored to neuromorphic computing architectures that employ the Differential Pair Integrator (DPI) circuit[13], [14] as a current-mode filter to emulate synaptic dynamics. The DPI circuit has a gain that depends on two independent parameters which can be used to represent a global synaptic scaling gain (e.g., controlled via a homeostatic control circuit), and a local synaptic weight (e.g., modified via local spike-based learning circuits).

This manuscript describes in detail the circuits for the gain control loop and for the ultra-low leakage CMOS cell that control the global synaptic scaling gain term. A detailed characterization performed on a test circuit implemented on a standard 0.18 µm CMOS process shows that the proposed circuit can modulate the total synaptic drive of a single neuron with extremely long time scales (up to 100 kilo-seconds) while maintaining unaltered the relative weight ratios among individual synapses obtained with spike-based plasticity circuits.

## II. THE HOMEOSTATIC AUTOMATIC GAIN CONTROL LOOP

Typical neuromorphic computing architectures comprise arrays of silicon neurons each receiving input from a large number of input synapses (see Fig. 1) [14], [16]. In these systems, it is possible to maintain neuron's overall spiking activity within given operating boundaries, without interfering with the network's signal processing and learning mechanisms, by adopting automatic gain control mechanisms that globally scale the synaptic weights of the synapse circuits afferent to their corresponding neuron over very long time periods. The DPI log domain integrator features two independently tunable parameters ($I_{\text{gain}}$ and $I_w$) for the control of the synaptic efficacy [13]. If all the synapses afferent to the neuron share the same temporal dynamics, it is possible to use one single integrator circuit per neuron with a single $I_{\text{gain}}$ parameter and use the temporal superposition principle to combine the output of multiple branches, with multiple independent synaptic plasticity parameters $I_{wi}$ that represent multiple synaptic inputs. It is therefore possible to control the global synaptic efficacy by using one single global scaling term, and multiple independent synaptic weight terms (e.g., see the multiple $I_{wi}$ currents in both Figs. 1 and 2). It has been shown that the circuit has first-order linear dynamics (see [14] for an analysis based on the translinear principle and [13] for a time-domain analysis).

In particular, if we consider the DPI circuit shown in the top part of Fig. 2, and assume that $\kappa_n = \kappa_p$ and $I_{syn} \gg I_{\text{gain}}$, then we can express its transfer function as:

$$\tau_s \frac{d}{dt} I_{syn} + I_{syn} = \frac{I_w I_{\text{gain}}}{I_\tau} \quad (1)$$

where the term $\tau_s$ is defined as $(C_{DPI} U_T)/(\kappa I_\tau)$, with $U_T$ representing the thermal voltage, and $\kappa$ the sub-threshold slope coefficient [17]. The current $I_\tau$ is a bias current that sets the integrator time constant. The current $I_w$ corresponds to the sum of the individual synapse currents $I_w = \sum_i I_{wi}$, set by their corresponding synaptic weight bias voltages $V_{wi}$. The current $I_{\text{gain}}$ represents an extra independent term which is defined as:

$$I_{\text{gain}} = I_0 e^{\frac{\kappa(V_{dd} - V_{THR})}{U_T}} \quad (2)$$

By adjusting the control voltage $V_{THR}$, the current $I_{\text{gain}}$ can be tuned so as to increase or decrease the steady state value of $I_{syn}$, independent of $I_w$ (which can be tuned by a regular learning process). A copy of the DPI output current $I_{syn}$ is eventually injected into the silicon neuron, which will then produce output spikes at a firing rate proportional to its amplitude. Fig. 2 shows how the full Automatic Gain Control (AGC) homeostatic control loop is used to modulate the voltage $V_{THR}$ in order to maintain the current $I_{syn}$ around a set reference current $I_{REF}$. In this loop, the $I_{syn}$ current is fed into a high-gain voltage comparator that compares the voltages that set the $I_{syn}$ and $I_{ref}$ currents. Depending on the outcome of this comparison, the output voltage $SW$ of this comparator is set to either ground or $V_{DD}$. This digital signal is then used to gate the control signals of a Low Leakage Cell (LLC) circuit which slowly adjusts $V_{THR}$ to up-regulate or down-regulate $I_{syn}$ accordingly.

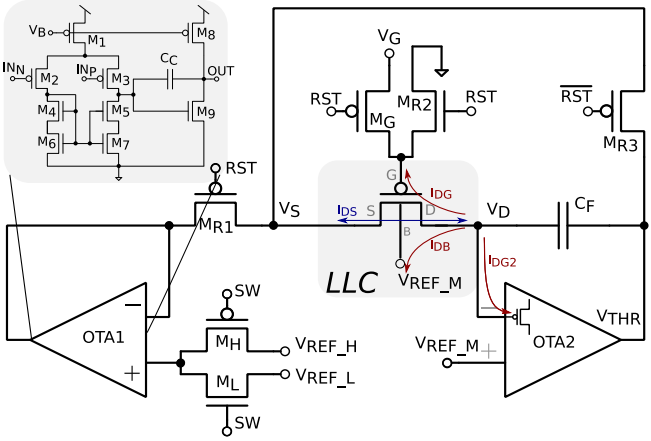

Fig. 3. Circuit implementation of the LLC used in the AGC loop.

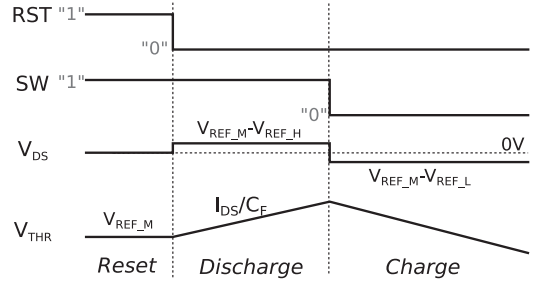

Fig. 4. Timing diagram of internal signals in AGC loop. RST is used to initialize the AGC loop to an initial condition. After resetting, $V_{DS}$ will be set to $V_{REF\_M} - V_{REF\_H}$ for $SW = 1$ and $V_{REF\_M} - V_{REF\_L}$ for $SW = 0$ to decide direction of leakage current $I_{DS}$ for discharging/charging integrate capacitor $C_F$. Slope of $V_{THR}$ is then $I_{DS}/C_F$.

### III. THE ULTRA-LOW LEAKAGE CELL

To achieve long biological realistic time-scales that do not interfere with signal transmission and learning, it is necessary to develop circuits with time scales that range from seconds to hours. In order to optimize the circuit's area to allow the dense integration of thousands of synapses and neurons on a single chip, the capacitance of the homeostatic control circuits must be small, therefore long time scales can only be achieved by extremely small currents. An example of such a circuit is the ultra-low leakage cell shown in Fig. 3. This circuit increases or decreases its output voltage $V_{THR}$ by controlling the direction of a very small current across the channel of the LLC p-FET to slowly charge or discharge the capacitor $C_F$. As in our LLC circuit implementation the capacitance $C_F$ is set to 1 pF, the currents required to obtain time scales in the order of thousands of seconds have to be of the order of 1 fA, which is usually one hundred times smaller than channel off-state leakage current of transistors, for a standard 0.18 $\mu$m CMOS process used.

Ultra-low ranges of currents can be obtained by minimizing the leakage currents across LLC p-FET with isolated N-well. In particular, the drain-to-bulk diode leakage current $I_{DB}$ of Fig. 3 can be minimized by biasing $V_{DB}$ to be zero [11]; this condition can be met by using a feedback Operational Transconductance Amplifier (OTA) with large enough gain (see OTA2 in Fig. 3). Minority carrier diffusion between the source and drain under the accumulation charge layer is reduced by applying very small $V_{DS}$ (normally several mV) across the p-FET. An isolated well for the p-FET is used to limit the number of junctions that can diffusively interact with drain-to-bulk junction. Also, a zero-bias ($V_B = V_D \approx V_S$) is used to minimized diffusive currents [11]. Furthermore, to get ultra-small leakage current from node D of the LLC p-FET, it is necessary to minimize the gate leakage currents $I_{DG}$ and $I_{DG2}$: the gate leakage current density normally is exponentially related to the thickness of gate oxide and strongly depends on gate bias [18]. For a standard 0.18 $\mu$m process with a gate oxide thickness of 4.6 nm, it is reasonable to assume that the gate leakage current density with gate bias of 0.5 V to be smaller than $10^{-8}$ A/m$^2$. To minimize these currents we designed the low leakage transistor with a $W/L$ ratio of 0.5 $\mu$m/1 $\mu$m and the p-FET input transistors of the OTA2 with a $W/L$ ratio of 8 $\mu$m/1 $\mu$m. Therefore the total gate leakage current is estimated to be smaller than 0.1 aA. By minimizing other leakage currents, a tunable subthreshold current $I_{DS}$ (tuned by $V_G$) is used to charge/discharge the capacitor $C_F$ While the OTA2 amplifier is used to implement a high-gain negative feedback loop to keep the potential of $V_D$ as close as possible to $V_{REF\_M}$, the OTA1 amplifier is used to clamp the voltage $V_S$ of the LLC p-FET to one of the two $V_{REF\_L}$, $V_{REF\_H}$ reference voltages. The detailed circuit schematic diagram of the OTA1 and OTA2 amplifiers is shown in the top-left inset of Fig. 3. To ensure high-gain and rail-to-rail output range, while minimizing power, we adopted a two stage pseudo-cascode split-transistor sub-threshold technique [19].

Fig. 4 shows signal waveform of proposed AGC loop. Given these small currents, At the beginning of an experiment it is necessary to quickly initialize the AGC control loop to a proper initial condition, such as $V_{THR} = V_D = V_{REF\_M}$. This can be done by enabling the digital control signal $RST$ to high, and resetting it to ground shortly after. At this point the direction of the current across the LLC p-FET of Fig. 3 will be set by the digital control signal $SW$, produced by the comparator of Fig. 2. If $SW$ is high, then the $V_{DS}$ of the LLC p-FET will correspond to $V_{REF\_M} - V_{REF\_L}$, otherwise it will correspond to $V_{REF\_M} - V_{REF\_H}$. By appropriately setting these reference voltages such that $V_{REF\_L} < V_{REF\_M} < V_{REF\_H}$, and by properly tuning the LLC p-FET's gate voltage $V_G$, it is possible to precisely control both direction and amplitude of the LLC p-FET $I_{DS}$ current.

The AGC loop of Fig. 3 implements a "bang-bang" control strategy: during normal operation, if the total synaptic drive $I_{syn}$ increases above the reference current $I_{REF}$, the comparator will set the digital signal SW to high. Since this enables the signal $V_{REF\_L}$ as input to the OTA1 amplifier, the current will slowly discharge $C_F$ and cause an increase in $V_{THR}$. This will in turn downscale the value of $I_{\text{gain}}$ of (2), effectively reducing the synaptic current $I_{syn}$ injected into neuron, and compensating for the initial change. Conversely, as $I_{syn}$ decreases below $I_{REF}$, the comparator will enable $V_{REF\_H}$ as input to OTA1. This will cause the LLC p-FET current to slowly charge the $C_F$ capacitor, thereby decreasing $V_{THR}$ and increasing $I_{\text{gain}}$. This will counteract the source of the disturbance that caused the initial decrease of $I_{syn}$, and increase it back, until it reaches again the reference level $I_{REF}$.

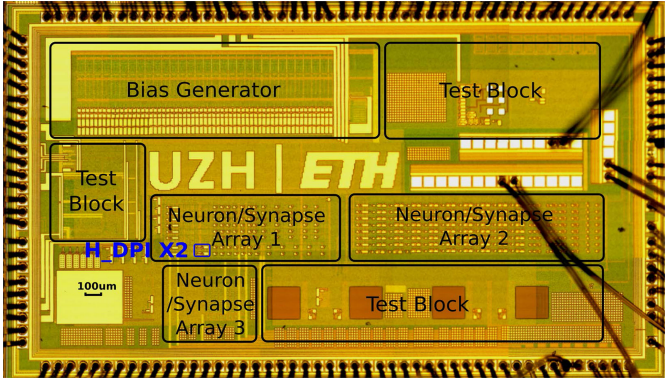

Fig. 5. Die photo of test chip implemented using a standard 0.18 $\mu$m CMOS process. The proposed DPI-based very long time scale automatic gain control synaptic scaling circuits are embedded in the Neuron/Synapse Array #1, and circled in blue. The whole chip occupies an area of 3.96 mm $\times$ 2.29 mm, and the synaptic scaling circuits occupy an area of 84 $\mu$m $\times$ 22 $\mu$m.

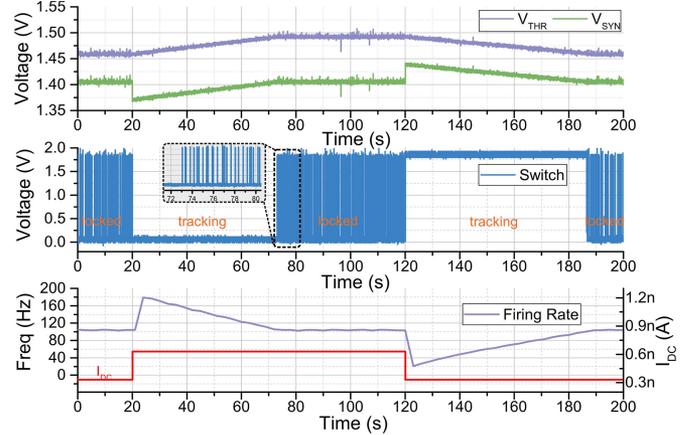

Fig. 6. Synaptic homeostasis measurements in response to step changes of the DPI input current. (Top): The voltage traces $V_{THR}$ and $V_{SYN}$; (Middle): the comparator output digital signal SW; (Bottom): Neuron's instantaneous firing rate and its input DC current.

## IV. Experimental Results

To characterize the response properties of the proposed circuits, we designed a prototype test chip in standard 0.18 $\mu$m CMOS process comprising a small array of neurons and synapses with embedded synaptic scaling circuits. Fig. 5 shows the die-photo of the fabricated chip, with the synaptic scaling circuits highlighted in neuron #1 (with its synaptic input array). An on-chip programmable bias generator [15] is implemented to generate all gate voltages.

In Fig. 6 we show the response of the circuit to a DC change in the input current $I_{DC}$ applied as synaptic weight input current into the circuit's DPI block (see also Fig. 2).

In this experiment we set $I_{DC}$ to start at 0.3 nA, the reference current $I_{REF}$ to be 20 nA, and the parameters of the silicon neuron (e.g., gain, time scales and refectory period) in a way to obtain a firing rate of approximately 100 Hz. $V_{REF\_H}$, $V_{REF\_M}$ and $V_{REF\_L}$ are set to 1.384 V, 1.382 V and 1.380 V, respectively. By setting the $V_G$ bias voltage of Fig. 3 to 1.42 V, we achieved adaptation time scales of approximately 60 seconds. In these conditions, the AGC loop of Fig. 2 clamps $V_{THR}$ to a value around 1.46 V, and $V_{SYN}$ around 1.4 V, thus maintaining the neuron's firing rate stable at its initial value. After 20 seconds $I_{DC}$ changes from 0.3 to 0.6 nA. As expected, this increased the DPI output current $I_{syn}$, decreased the $V_{SYN}$ voltage accordingly, and increased the neuron's firing rate from 100 to about 180 Hz. The synaptic scaling homeostatic circuits turn on and slowly scale down the total synaptic current $I_{syn}$ being injected in the neuron, which in turn starts to slowly decrease its output firing rate. This is done by increasing the $V_{THR}$ signal, which is shared by all input synapses afferent to the same neuron, and which modulates the $I_{\text{gain}}$ current. After approximately 60 seconds $I_{syn}$ and the firing rate of the neuron are both restored to their initial values. At around $t = 120$ s $I_{DC}$ changes back from 0.6 to 0.3 nA. In this case, the neuron's firing rate drops below its original value and the AGC loop is activated in the opposite direction, such that after about 60 seconds, the neuron's firing rate is restored back to its original value. Due to the *bang-bang* nature of the AGC control loop, when the neuron's firing rate is close to the reference the homeostatic circuits keep on alternating the SW signal from high to low, in order to keep the $I_{syn}$ current around the $I_{REF}$ reference current (see "locked" regions in Fig. 6). The continuous switching of SW in the "locked" regions is kept at slow rates of several Hz to tens Hz (see zoomed-in area in the inset of Fig. 6) by limiting the feedback current flowing through LLC to charge/discharge the capacitor $C_F$. The continuous switch frequency of SW during "locked" regions is normally very slow (from several Hz to tens Hz) because of the ultra low loop bandwidth. This results in very low power consumption, also in this condition.

Figs. 7 and 8 show the effect of changing the recovery rate, that can be achieved by appropriately changing the $V_G$ bias voltage of LLC p-FET, which sets the amplitude of the $I_{DS}$ current on Fig. 3, and by modulating the difference between $V_{REF\_M}$, and $V_{REF\_L}/V_{REF\_H}$, which control the voltage drop across the LLC p-FET channel. In Fig. 7, the target neuron is initially stable at a firing rate of 150 Hz and is abruptly changed to 475 Hz (at $t = 60$ s) by applying a step current of 2.5 nA. The step change in firing rate activates the homeostatic mechanism that slowly adapts the neuron's output to recover the original activity, with time scale of 75 s and 150 s, respectively.

Fig. 8 shows that it is possible to get longer time constants and that the control is symmetrical, working both for increase and decrease of the neuron's firing rate away from its equilibrium.

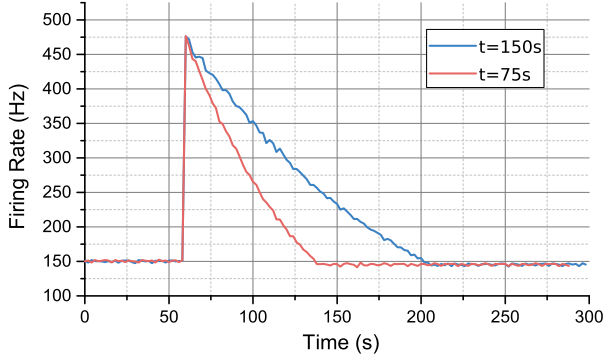

Fig. 7. Neuron's firing rate modulated by the homeostatic mechanism with different time scales.

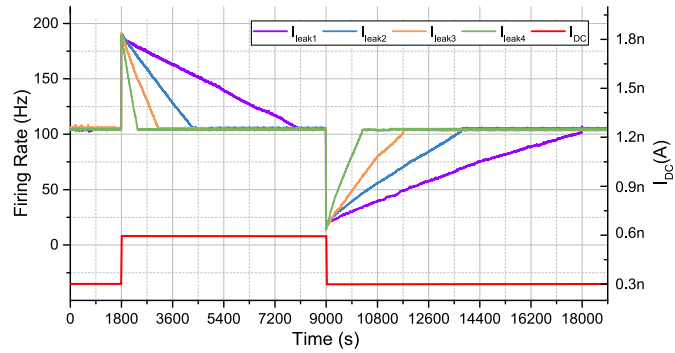

Fig. 8. Neuron's firing rate modulated by the homeostatic mechanism, tuned to respond with different time scales. The bottom red curve represents the DPI's input current $I_{DC}$.

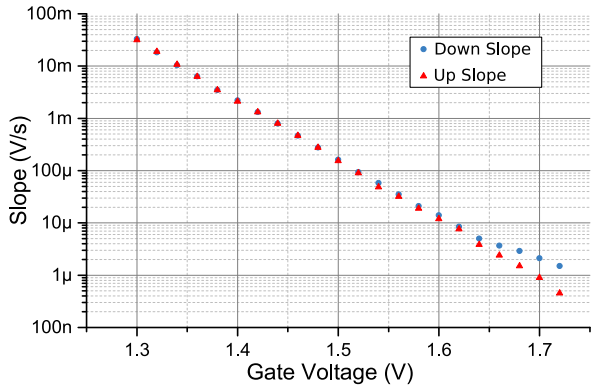

Fig. 9. Measured charge/discharge slope of $V_{THR}$ with controllable leakage current from LLC. $V_{REF\_H}$, $V_{REF\_M}$ and $V_{REF\_L}$ are set to 1.384 V, 1.382 V and 1.380 V, respectively, and $V_G$ is swept from 1.3 V to 1.72 V.

In each condition the AGC succeeds in restoring the neuron's activity to the 100 Hz rate defined as target point.

In order to exploit the achievable time scales of proposed AGC, we measured slope of $V_{THR}$ when charging/discharging the 1 pF $C_F$ with controllable leakage from LLC. In this experiment, $V_{REF\_H}$, $V_{REF\_M}$ and $V_{REF\_L}$ are set to 1.384 V, 1.382 V and 1.380 V, respectively, and $V_G$ is swept from 1.3 V to 1.72 V. As is shown in Fig. 9, Up/Down slope decays exponentially for an increasing $V_G$. For a $V_G > 1.5$ V, Up/Down slope starts to be different for a discharge/charge leakage current smaller than 100 aA. The slope will be significantly different for $V_G > 1.62$ V when the discharge/charge leakage current is smaller than 10 aA. This asymmetrical maybe caused by diffusion current between the drain-to-bulk diode and leakage current from gate leakage as analysed in Section III. At $V_G = 1.72$ V, the Up/Down Slope is 1.5 $\mu$V/s and 0.45 $\mu$V/s with a leakage current of 1.5 aA and 0.45 aA, equivalent to 9.4 Electrons/second and 2.8 Electrons/second, respectively.

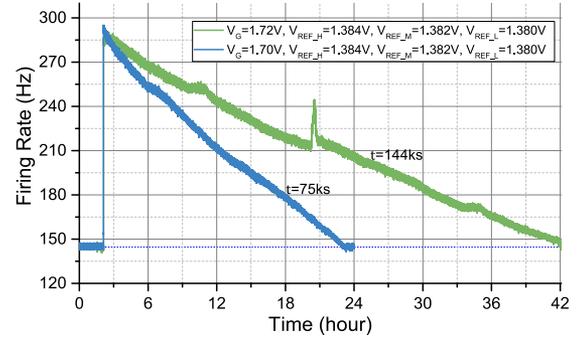

Fig. 10. Neuron's firing rate modulated by the homeostatic mechanism, tuned to produce the longest time scales.

Fig. 10 shows the response of neuron's firing rate with time scales of around 75 ks and 144 ks. In this experiment, $V_{REF\_H}$, $V_{REF\_M}$ and $V_{REF\_L}$ are set to 1.384 V, 1.382 V and 1.380 V, respectively. $V_G$ is tuned to be 1.7 V and 1.72 V for achieving these long time scales. The observed peak on neuron's firing rate curve with time scale of 144 ks is caused by sudden temperature drifts at noon during the experiment which led to transient changes of biases generated by on-chip bias generator [15].

While the first characterization of the homeostatic control was obtained by artificially changing the neuron drive with an externally injected current, Figs. 11 and 12 show that the control is effective when the total neuron drive changes after its synaptic drive changes due to the effect of spike-based learning, which leads to synaptic potentiation. The homeostatic plasticity mechanism employed in this context is useful for avoiding the risk of runaway potentiation.

In Fig. 11, we show a neuron that receives input on six afferent synapses, which feed currents to the target neuron that are proportional to their synaptic weights. The analog weights of the synapses are represented by $I_{wi}$ The synapse circuits include spike-based learning mechanisms that can potentiate the synapse (i.e., increase it's $I_{wi}$) or depress it, depending if input and output firing rates are correlated or not (see [16] for a detailed description of the learning circuits and behavior). We drive the neuron with an external signal and stimulate the synapses with input signals in a way to trigger the learning mechanism to potentiate or depress the weights of the synapses being stimulates. Specifically, at the beginning of the experiments we set all synapses to the depressed state, and we provide a teacher current to the neuron to maintain its firing rate to around 80 Hz. Fig. 11(a) shows what happens when at time $t_1 = 70$ s, $t_2 = 105$ s and $t_3 = 140$ s, 2/4/6 synapses are sequentially potentiated, and then stimulated by a spike train with Poisson distribution around a mean rate of 100 Hz. Eventually, the weights of all synapses are reset to low at time $t_4$. Fig. 11(b) shows the response of the neuron, that adjusts its firing rate thanks to the homeostatic control that acts on $I_{\text{gain}}$. Initially $I_{\text{gain}}$ is big and the effect of a single input spike is high on the neuron's membrane potential. The potentiation of the synapses has a first effect of changing the mean output firing rate of the neuron for the same input spike train, however, the homeostatic mechanism decreases $I_{\text{gain}}$ such that the effect of a single spike provokes a smaller change on the membrane potential of the neuron. Fig. 12 confirms the qualitative behavior of the control with a different number of active synapses.

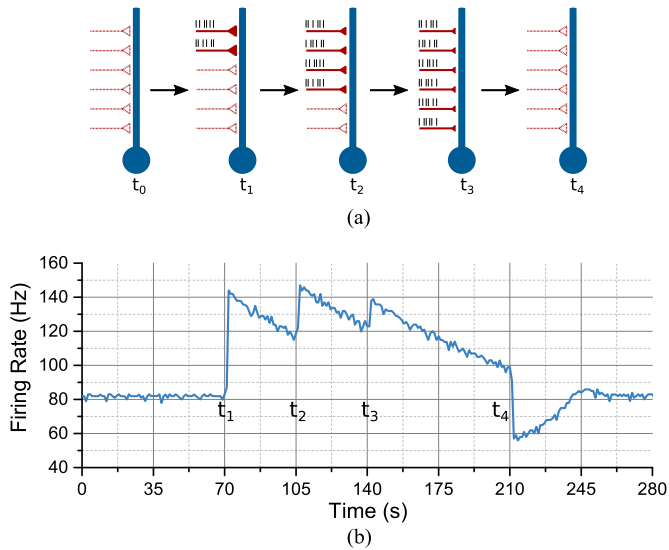

Fig. 11. Homeostatic control effect after spike-based synaptic potentiation. (a) Initially all synaptic weights are set to non-potentiated, at times $t_{1,2,3}$ two synapses are potentiated, while applying poisson input events with mean rate of 100 Hz. And weights of all synapses are reset to low at time $t_4$. (b) A teacher current is injected to neuron to maintain its firing rate to around 80 Hz, the firing rate is then changed by the synaptic input and restored to its initial level by the homeostatic control.

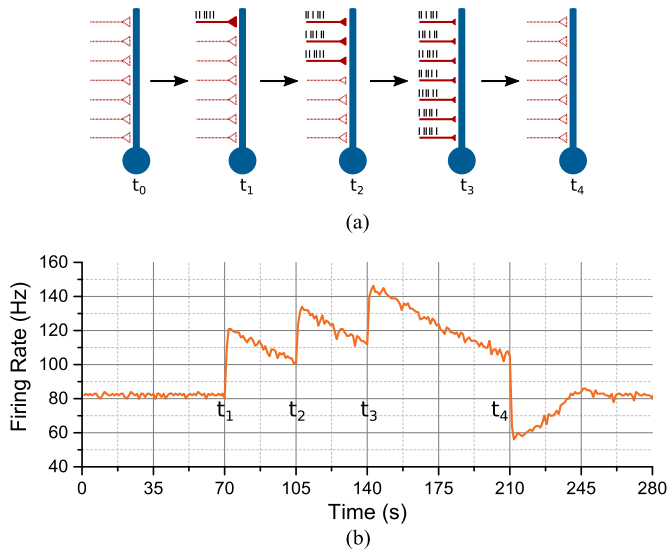

Fig. 12. Homeostatic control effect after spike-based synaptic potentiation. The experiment is the same as in Fig. 11 but for a different number of synaptic inputs.

TABLE I
CURRENT LEAKAGE COMPARISON

|  | [10] | This work |
|---|---|---|
| Technology | 0.18 $\mu$m | 0.18 $\mu$m |
| Power supply | 1.8 V | 1.8 V |
| Area | 83 $\mu$m × 42 $\mu$m | 84 $\mu$m × 22 $\mu$m |
| Power consumption | 100 nW | 10.8 nW |
| Time scale | ms-400 s | ms-144 ks |
| Leakage slope (1 pF) | - | 1.5 $\mu$V/s / 0.45 $\mu$V/s |
| Leakage | 210 aA | 1.5 aA / 0.45 aA |
| Electrons per second | 131 $e^-$/s | 4.9 $e^-$/s / 2.8 $e^-$/s |

Table I shows a direct comparison of the proposed homeostatic plasticity circuit with state of the art circuits described in the introduction in terms of current leakage performance and general circuit performance indicators, respectively, showing that the proposed implementation outperforms them in circuit area, power consumption and leakage current, achieving the highest possible time constants.

## V. CONCLUSIONS

In our previous work, we presented the DPI as a neuromorphic synapse circuit that implements biologically realistic synaptic dynamics while supporting both spike-driven learning mechanisms and synaptic scaling homeostatic plasticity mechanisms [13], [14]. In this paper, we proposed a homeostatic plasticity circuit that exploits the features of the DPI to implement synaptic scaling homeostasis. The circuit we proposed here globally scales the weights of all synapses impinging on the same post-synaptic neuron, with time scales that can range from milliseconds to days. The ability to globally scale synaptic weights on very long time scales and without affecting with their ratios is important, as it allows the system to compensate for global and slow changes in both the input signals and the system properties without interfering with the spike-based learning mechanisms that change the weights depending on the statistics of the input signals. However, the ability to precisely control the temporal scale of the homeostatic process, and to set faster time constants in the homeostatic control loop is also important, as it has been recently shown that the interaction of these types of homeostatic processes with the conventional learning mechanisms produce hetero-synaptic competition that improves the ability of the network to generalize and to maximize its memory storage capacity [1], [7].

To validate the proposed circuit design, we fabricated and tested an ultra-low leakage cell that allowed us to obtain extremely long time constants in a controllable way. We measured the low leakage currents obtained from well-biased single p-FET device and demonstrated how, with a 1 pF capacitor, it is possible to control leakage currents as small as 0.45 aA (i.e., less than 3 electrons per second) and reach time scales as large as 144 k seconds (i.e., more than 40 hours). The proposed circuits occupies an area of 84 $\mu$m × 22 $\mu$m in a standard 0.18 $\mu$m process, and consumes 10.8 nW with 1.8 V supply power during normal operation. In comparison to previously proposed designs, this circuit does not require additional floating gate devices or off-chip methods. This makes it suitable for dense integration with other low-power neuromorphic circuits in the next generation of neuromorphic computing platforms.

## ACKNOWLEDGMENT

The authors would like to thank the Reviewers who gave extremely valuable feedback and did an excellent job at analyzing the system details (from the model equations to the circuits), pointing out important details, that have now been incorporated into the manuscript.

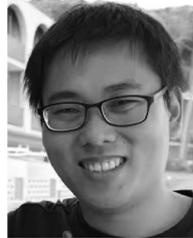

**Ning Qiao** (M'15) received the Bachelor's degree in microelectronics and solid-state electronics from Xi'an Jiaotong University, Xi'an, China, in 2006, and the Ph.D. degree in microelectronics from the Institute of Semiconductors, Chinese Academy of Sciences, Beijing, China, in 2012, researching on ultralow-power low-noise mixed-signal circuits in SOI process. He is a Postdoctorate Researcher in the Institute of Neuroinformatics, University of Zurich and ETH Zurich, Zurich, Switzerland, focusing on developing mixed-signal multicore neuromorphic VLSI circuits and systems. His current research interests include ultralow-power subthreshold mixed-signal neuromorphic VLSI circuits and systems, parallel neuromorphic computing architectures, and fully asynchronous event-driven computing/communication circuits and systems.

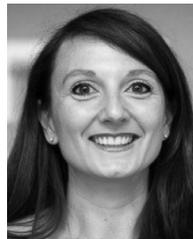

**Chiara Bartolozzi** (M'12) received the degree in engineering (with Hons.) from the University of Genova, Genova, Italy, and the Ph.D. degree in neuroinformatics from ETH Zurich, Zurich, Switzerland, developing analog subthreshold circuits for emulating biophysical neuronal properties onto silicon and modeling selective attention on hierarchical multichip systems. She is a Researcher in the Istituto Italiano di Tecnologia, Genova, Italy. She is currently leading the Event Driven Perception for Robotics Group, mainly working on the application of the "neuromorphic" engineering approach to the design of sensors and algorithms for robotic perception.

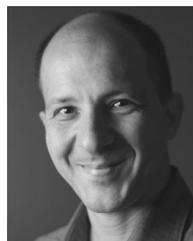

**Giacomo Indiveri** (SM'06) is a Professor at the Faculty of Science of the University of Zurich, Switzerland, and director of the Institute of Neuroinformatics of the University of Zurich and ETH Zurich. He received the M.Sc. degree in electrical engineering and the Ph.D. degree in computer science from the University of Genoa, Italy. He was a post-doctoral research fellow in the Division of Biology at Caltech and at the Institute of Neuroinformatics of the University of Zurich and ETH Zurich. He holds a "habilitation" in Neuromorphic Engineering at the ETH Zurich Department of Information Technology and Electrical Engineering. He was awarded an ERC Starting Grant on "Neuromorphic processors" in 2011 and an ERC Consolidator Grant on neuromorphic cognitive agents in 2016. His research interests lie in the study of neural computation, with a particular focus on spike-based learning and selective attention mechanisms. His research and development activities focus on the full custom hardware implementation of real-time sensory-motor systems using analog/digital neuromorphic circuits and emerging VLSI technologies.